\documentclass[conference]{IEEEtran}
\usepackage{cite}
\usepackage{amsmath,amssymb,amsfonts}
\usepackage{algpseudocode}
\usepackage{graphicx}
\usepackage{textcomp}
\usepackage{tabularx}
\usepackage{booktabs}
\usepackage{xcolor}
\usepackage[hidelinks]{hyperref}
\usepackage{verbatim}
\usepackage{makecell}
\usepackage{algorithm}
\usepackage{threeparttable}
\def\BibTeX{{\rm B\kern-.05em{\sc i\kern-.025em b}\kern-.08em
    T\kern-.1667em\lower.7ex\hbox{E}\kern-.125emX}}
\begin{document}
\title{
Fully Decentralized and Safety-Aware Multi-Agent Reinforcement Learning for Control on Networks 
}
\author{\IEEEauthorblockN{Theodore Rogalski}
\IEEEauthorblockA{\textit{Electrical \& Computer Engineering Department} \\
\textit{Stevens Institute of Technology}\\
Hoboken, USA \\
trogalsk@stevens.edu}
\and
\IEEEauthorblockN{Shirantha Welikala}
\IEEEauthorblockA{\textit{Electrical \& Computer Engineering Department} \\
\textit{Stevens Institute of Technology}\\
Hoboken, USA \\
swelikal@stevens.edu}}
\maketitle
\begin{abstract}
This paper develops a safe and fully decentralized multi-agent reinforcement learning (MARL) algorithm to solve a class of discrete-time control problems on networks, including the persistent monitoring problem. 
Fully decentralized control of agents, while offering numerous benefits, faces issues such as exponentially increasing sample complexity, lack of global information about the system, and challenges in coordinating between agents. 
To address these issues, this paper introduces a fully decentralized multi-agent reinforcement learning algorithm that integrates deep reinforcement learning with safety considerations. This method feeds a history of local observations of the network’s state into two parallel neural-network branches: the graph encoder, which adds structural information and correlations among nodes, and a state estimator, which predicts the uncertainty at each node in the graph. Additionally, the result of feeding that input into an actor-critic network is passed through a discrete-time control barrier heuristic to reduce the likelihood that any node will be neglected.
This approach enables teams of fully decentralized agents to solve challenging problems by increasing system awareness and incorporating built-in safety measures to prevent the adoption of potentially harmful control policies.
Numerical results from a custom simulation environment demonstrate that the proposed algorithm achieves 26.3\% lower average uncertainty than a centralized control policy and is within 1\% of the uncertainty performance of a more computationally complex algorithm with added attention layers.
\end{abstract}

\section{Introduction}
\label{sec: introduction}

Multi-agent reinforcement learning (MARL) \cite{amato2025initialintroductioncooperativemultiagent} is a powerful method for solving discrete-time control problems autonomously, but current approaches rely on unrealistic assumptions such as perfect state information and various forms of communication among agents. Those centralized MARL approaches contrast with fully decentralized MARL, an approach that replaces a single MARL controller that has access to full-state information and action selections for each agent with independent controllers with no information sharing during training or evaluation, including each agent's local representation of the state, network parameters, and all other pieces of data. Fully decentralized approaches remove the overhead of communicating and sharing parameters with other agents but make coordination more difficult and lack convergence guarantees.
Replacing the centralized controller with independent controllers for each agent presents three  main issues: the lack of global state renders traditional methods for ensuring convergence of the algorithm to an optimal policy moot, the fact that each controller is independent removes inherent coordination from the architecture, and the curse of multi-agency leads to exponential sample complexity with respect to the number of agents for fully decentralized algorithms \cite{shi2025breakingcursemultiagencyrobust}. 

To address these challenges, decentralized algorithms may utilize inter-agent communication \cite{Lu_Zhang_Chen_Basar_Horesh_2021,oh2025consensusbaseddecentralizedmultiagentreinforcement,zhang2025finitetime,yuan2015convergencedecentralizedgradientdescent}, but its use incurs computational and implementation costs. Deep reinforcement learning (DRL) \cite{deeprlsurvey}, an approach that incorporates deep learning components, such as neural networks, into reinforcement learning algorithms, enables strong generalization without requiring comprehensive system information \cite{zhangcompass}.

Another decentralized method for addressing those challenges is safe reinforcement learning, referred to as "safety" in this work. Complementary to DRL, safety methods \cite{bejarano2024safety,gu2024reviewsafereinforcementlearning} enable deterministic behavior design and agent coordination in control architectures, even without the ability to communicate or share information. The stochastic nature of DRL is augmented by the predictable nature of safety methods, which we propose may increase the reliability of fully decentralized control policies. However, even with the added reasoning ability of DRL and safety integration, agents optimize for local sub-goals rather than the global picture, necessitating the selection of a problem whose solution fits that paradigm.

The persistent monitoring problem is a problem formulation in which a team of agents must be coordinated to minimize a global measure of node uncertainty by monitoring the state of certain valuable points of interest \cite{guopersistentmonitoringforpointsofinterest}. Persistent monitoring is uniquely suited to a solution that optimizes local sub-goals, as it covers a large area that agents can sample and monitor efficiently. However, a realistic persistent monitoring scenario involves a large number of system variables that are abstracted away by the model, such as electronic noise and physical constraints. This motivates representing the environment as a network of nodes and edges, with points of interest mapped to nodes and important spatial information mapped to edges between nodes. Each agent is allowed to move along these edges to these nodes, preserving the important connections between points in space while reducing unnecessary computation. 

This paper is organized as follows. Section \ref{sec: problem formulation} discusses the particulars of controlling a decentralized team of agents and the selected persistent monitoring problem as the specific control problem to address. The proposed algorithm architecture is discussed in Section \ref{sec: methodology}, with supporting experiments and results detailed in Section \ref{sec: numerical results}.



\section{Problem Formulation}
\label{sec: problem formulation}

\subsection{Control of Agents on Networks}
\label{sub: control of agents on networks}
We consider a dynamic environment with decision-making agents $\mathcal{A}\triangleq\{\mathcal{A}_i:i\in\mathbb{N}_N\}$ (where $\mathbb{N}_N\triangleq\{1,2,...,N\}$) that can be modeled as a network $\mathcal{G}\triangleq(\mathcal{V},\mathcal{E})$. The nodes $\mathcal{V}\triangleq\{\mathcal{V}_m:m\in\mathbb{N}_M\}$ and the edges $\mathcal{E}\subset \mathcal{V}\times\mathcal{V}$ represent entities of interest and their interconnections within this environment. In addition, each node $\mathcal{V}_m$ may have features $\mathcal{F}_m\in\mathbb{R}^n$, which can respond to agent behavior around the node. At the $k$\textsuperscript{th} time-step $k\in\mathbb{N}_K$, agent $\mathcal{A}_i$ makes an action selection $a_{i,k}\in A_{i,k}$ where $A_{i,k}$ is the local action space which is dependent on its local observations $\Omega_{i,k}$ and $\mathcal{A}_i$'s observation at $k$ of its one-step neighborhood $\mathcal{N}_{i,k}$ alongside its current position $\phi_{i,k} \in \mathcal{V}$, including edges and nodes $\Omega_{i,k} \triangleq \{\mathcal{N}_{i,k},\phi_{i,k}\}$. At each time-step $k$, an agent $\mathcal{A}_i$ may choose to move to a node directly connected to its current node $\phi_{i,k}$ or remain at $\phi_{i,k}$. 

Each node $\mathcal{V}_m$ is associated with a feature vector $\mathcal{F}_{m,k}$. These features may be used to model arbitrarily complex systems with arbitrarily complex dynamics; for example, a team of autonomous robots operating within a rugged three-dimensional space may be modeled as a network whose nodes and edges represent important spatial information. Fitting engineering problems to this formulation reduces irrelevant complexities; in this case, the persistent monitoring problem is abstracted as a representative example of complex problems.

\subsection{Persistent Monitoring}
The persistent monitoring problem, wherein a team of one or more agents is controlled to minimize uncertainty about an environment (mission space), has many characteristics that naturally allow it to be modeled using the above formulation, particularly because the node-edge connection structure lends itself to representing points of interest. The persistent monitoring problem serves as a strong representative for complex problems, meaning that a strong solution to it suggests applicability to other,t similar problems.

\begin{figure}
    \includegraphics[width=1\linewidth]{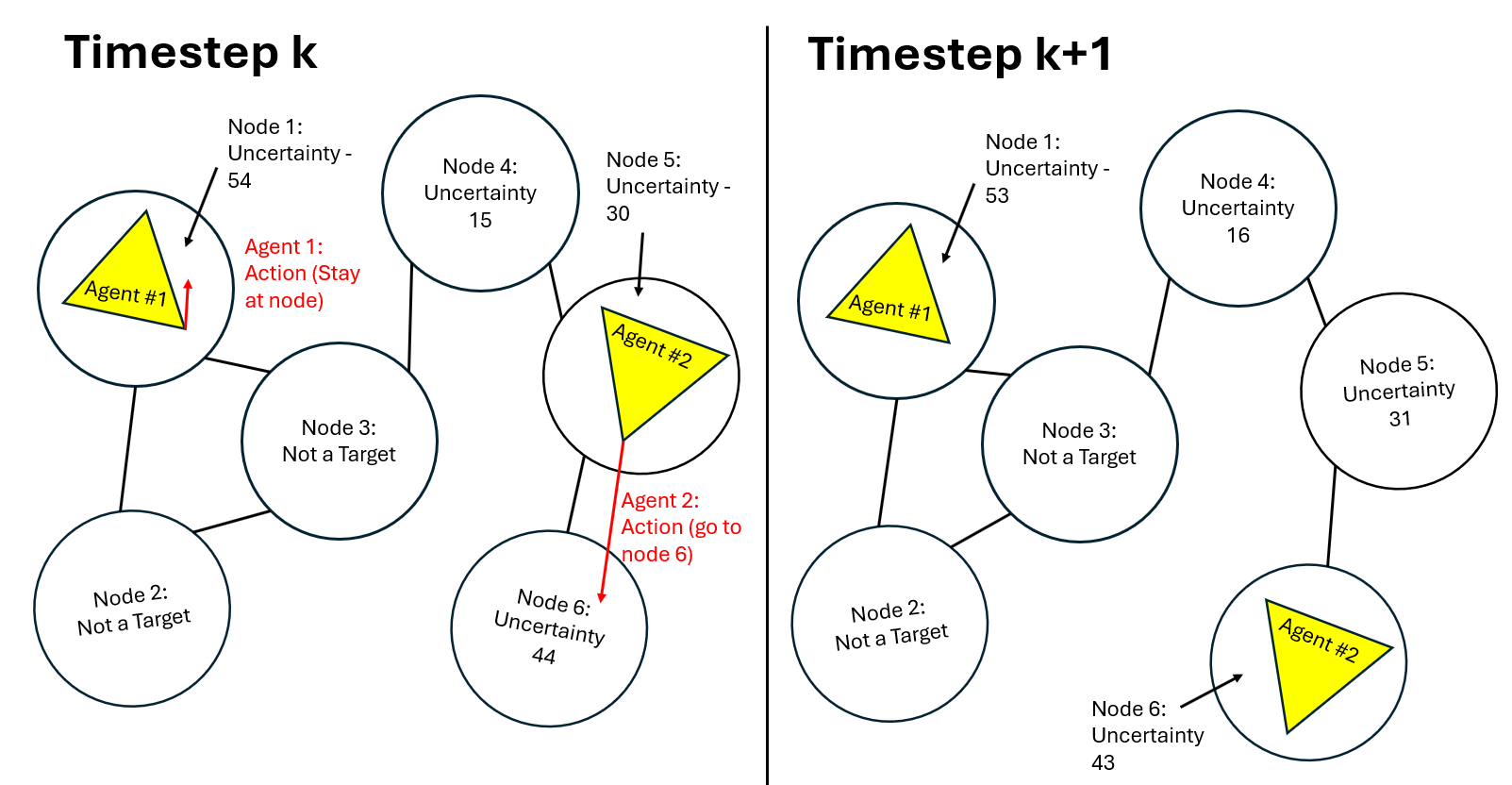}
    \caption{A persistent monitoring system with six nodes and two agents at discrete time steps $k$ and $k+1$.}
    \label{fig:persistentmonitoring}
\end{figure}

Applications of persistent monitoring include the monitoring of a physical area with ground-based robots \cite{Smith_2012} as well as the autonomous coordination of surveillance drones \cite{boldrer2026aerialrobotspersistentmonitoring}, a burgeoning field with various applications including in agricultural, military, and urban surveillance \cite{kimarriseofuav}. 
In order to solve the persistent monitoring problem using simulation, an accurate and abstract formulation must first be established: in the network $\mathcal{G}$, certain nodes $\mathcal{T}$ (targets) on the network $\mathcal{G}\triangleq(\mathcal{V},\mathcal{E})$ have uncertainty values (e.g., $u_m$ for a target node $\mathcal{V}_m$) that increase when they are not being monitored and decrease when they are monitored (i.e., visited 
 ) such that the uncertainty $u_{m,k+1}$ at a time-step $k+1$ is given by
\begin{equation}
u_{m,k+1}=
\begin{cases}
0, & t_m=0, \\
u_{m, k}+(1-2p_{m,k}), & t_m=1,u_{m, k}>0,\\
u_{m, k}+(1-p_{m,k}), & t_m=1,u_{m, k}=0,\\
\end{cases}
\end{equation} 
where $p_{m,k}\in\{0,1\}$ denotes the presence of an agent at node $\mathcal{V}_m$ at time-step $k$ and $\mathcal{t}_m\in\{0,1\}$ denotes whether or not each node is a target. \ref{fig:persistentmonitoring} shows an example of a persistent monitoring system with six nodes and two agents transitioning from time-step $k$ to $k+1$. 

\section{Methodology}
\label{sec: methodology}
\subsection{Algorithm Architecture}

\begin{figure}[!t]
\centering
\includegraphics[width=0.5\textwidth]{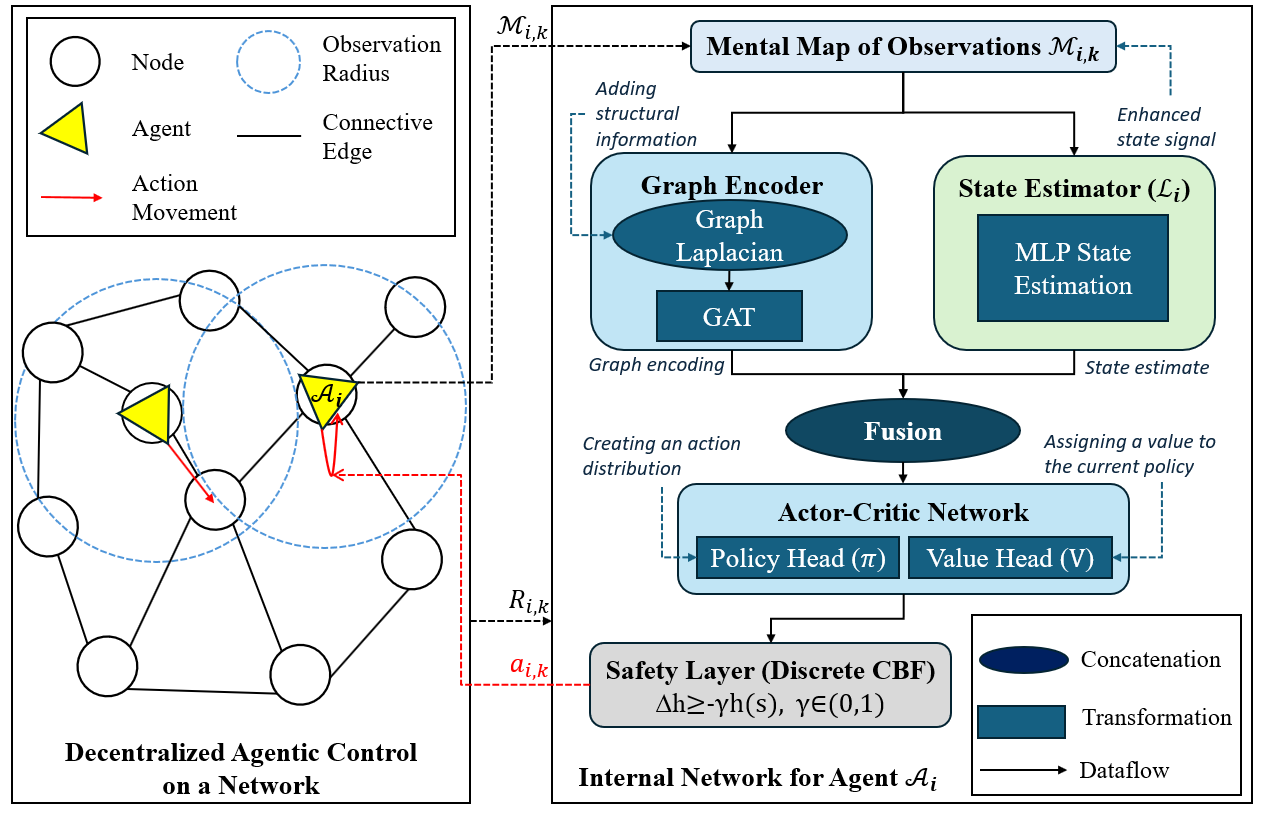}
\caption{At time-step $k$, the agent $\mathcal{A}_i$ receives $\mathcal{M}_{i,k}$, which is processed in the following manner: concatenation with $\mathcal{L}_i$, transformation by Graph Attention (GAT), concatenation with $\mathcal{L}_i(\mathcal{M}_{i,k})$ and then transformation by the Actor network. The result of the processed observation is passed through a safety layer and then sent to $\mathcal{A}_i$ as a probability mass function over possible actions.}
\label{fig:model_architecture}
\end{figure}

To address the persistent monitoring problem using decentralized agents, this project developed an algorithm that incorporates the following deep learning components: a lightweight mental map architecture and a discrete control barrier function (DCBF) heuristic, combined with existing deep reinforcement learning (DRL) algorithms. Each agent constructs an input to its deep network by concatenating the observations it has received, which is then augmented with the graph Laplacian. In parallel, its observation is sent to a neural network state estimator, which predicts the amount of uncertainty at each node at each time step. Then, the output from a Graph Attention network, concatenated with the state estimator result, is fed to an actor-critic network. The actor's output (the policy $\pi(a|s)$) is then fed into the DCBF heuristic, which removes the probabilities of agents taking actions that move to lower-uncertainty nodes. 

\subsection{Mental Map} Firstly, each agent $\mathcal{A}_i$ is given a mental map $\mathcal{M}_{i,k}$ of the network to allow for reasoning over time. If each  $\mathcal{A}_i$ has a potentially limited ability to observe the network $\mathcal{G}$ at a time step $k$ (i.e., $\Omega_{i,k}\subset\mathcal{G}$), then its state signal will be an incomplete representation of the network. To address incomplete system information, at each time step the mental map is updated with the latest local observation: newly observed features replace their stored values, while unobserved features retain their previous values. The mental map is initialized with empty nodes (edges are added during training/test time) to enable statically sized neural networks, saving computational bandwidth.

\subsection{Graph Laplacian} Then, the agent's transformed state signal is augmented with the graph Laplacian of $M_{i,k}, L_{i,k}=D_{i,k}-A_{i,k}$, where $D$ is the degree matrix and $A$ is the adjacency matrix of the mental map at $k$. The concatenation of $L$ to $M$ allows the agent to have built-in structural information about the graph, enabling it to move between nodes without having full information.
\subsection{Graph Attention} After receiving the enhanced state signal $\mathcal{M}_{i,k}$ concatenated with $L$, $\mathcal{A}_i$ reasons over it using the graph attention mechanism to learn correlations among neighboring nodes. Graph attention \cite{velickovic2018graphattentionnetworks} is defined as the following 
\begin{equation}
	\vec{f}'_m = \parallel_{h=1}^H \sigma\Big(\sum_{j\in\mathcal{N}_m}\alpha_{mj}^h{\bf W}^h\vec{f}_j\Big),
\end{equation}
where $f'_m$ are the transformed features of $\mathcal{V}_m$ $\mathcal{F}_m$, $\sigma$ is a non-linear function (in this case, LeakyReLU), $\mathcal{N}_m$ is each node connected to $\mathcal{V}_m$ via an edge, $\alpha_{m,j}$ is a softmax over the result of an attention mechanism between neighboring node $v_j$ and the node $u_m$, $\textbf{W}^k$ and $h\in H$ are attention heads that learn different associations between the features of $\mathcal{V}_m$ and that of surrounding nodes.
\subsection{State Estimator} At any given moment, the total ground-truth information an agent has about the system is limited to a composition of local observations of nodes and edges $\Omega_{i}$. Given that a subset of the network may not be visible to each agent due to that limited capacity to observe the environment, the estimation of such invisible node features $\mathcal{F}_m, \forall \mathcal{V}_m \in \mathcal{G}-\Omega_{i,k}$ is an important way for the agent to reason over the graph. At each time step, $\mathcal{A}_i$ uses its own Multilayer Perceptron (MLP) state estimator $\mathcal{L}_i$ for this estimation task. Each state estimator is updated according to the mean-squared prediction loss:
\begin{equation}
    \text{loss}(\mathcal{L}_i)=\frac{1}{|\mathcal{V}|}(\mathcal{L}_i(\mathcal{M}_{i,k-1})-\mathcal{M}_{i,k})^2
\end{equation}
where $s$ and $s'$ are the current and next states, respectively. $\mathcal{L}$ allows the agent to learn a mapping between the dynamics of any given state to the next state over time. Because $\mathcal{L}$ is a deep neural network, similar states may be mapped to similar outputs, allowing for a degree of generalization
\cite{oh2025consensusbaseddecentralizedmultiagentreinforcement}, \cite{fangweihaorlaircraft}. In previous works, communication assumptions eliminated the property of full decentralization from these algorithms, but the ability to guide training remains very useful in a fully decentralized setting.  After the graph encoder and state estimator representations are concatenated, the data is fed into a linear policy head to learn a complex environment. However, the fitness of $\pi$ is not available to the actor, so a learned value function $v_\pi(\mathcal{M})$, implemented as a multi-layer perceptron, is used to evaluate the current policy and improve it during training. Feeding the concatenation of the graph encoder and graph Laplacian to the actor-critic layer allows the agent to learn control policies faster, reducing the training time required to achieve good results and addressing the curse of multi-agency. It is implemented using two linear models: the policy head has input dimensions of 5*number of nodes, and the critic is a linear model with 3*number of nodes and 1 output layer for probabilities. The use of linear layers keeps computational cost low while allowing for learnable weights.

\subsection{Discrete-Time Control Barrier Function} The use of a Discrete-Time Control Barrier Function (DCBF) heuristic to ensure agents select actions that preserve safety, in this case, selecting nodes with the highest uncertainty, allows for greater control over the learned policy.

 Safety is defined as ensuring that no target exceeds a certain uncertainty threshold (set here to $k$), with the aim of reducing the number of unattended nodes. The safe set of actions $\mathcal{S}_{i,k}$ represents all actions for agent $i$ at time step $k$ that do not violate the principle of reducing uncertainty to a preset level $\gamma\in\mathbb{R}_{\geq0}$. The prediction of all uncertainties $\hat{u}_k$ at time step $k$ for the mental map $\mathcal{M}$ is used as the estimate of the state.

The safety value $h_{i,k}=\gamma_k-\text{max}(\hat{u}_k)$, where $\gamma_k$ represents a threshold after which each node must be attended to by an agent, given that the agent can estimate the uncertainty to be at that level; this threshold is equal to the number of time-steps during training and testing. Given these variables, the following process is taken to transform the probability distribution outputted by $\pi(a | s)$ to the safe distribution:
\subsubsection{Masking of Nodes} The selection of an action by an agent is limited to the set of all nodes on or one step away from its current node; the action at $k$ is $a_k\in\{\mathcal{N}_{i,k},\phi_{i,k}\}$. Therefore, all nodes not in this set are unreachable, and the probability of moving to them is set to 0.

\subsubsection{Lower Bound Computation} 
First, the predicted uncertainty $\mathcal{L}(\mathcal{V}_m)$ of each node is calculated.
Then, the maximum of these expected values is multiplied by $1-\eta$, $\eta\in[0,1]$ ($\eta$ is chosen to be 0.1) to ensure that any nodes whose uncertainty is not at least within 0.9 of the max uncertainty are moved to. 
\begin{equation}
    h_t=\text{max}(\mathcal{L}(\mathcal{V}_m),\mathcal{V}_m\in{\mathcal{V}})(1-\eta)
\end{equation}
\subsubsection{Unsafe Node Determination}
After $h_t$ is determined, each prediction $\hat{u}_{i,m,k}$ in the graph is looped through, checking if its uncertainty prediction is greater than or equal to $h_t$; if it is, its probability of being selected as an action is multiplied by one; otherwise, it is multiplied by zero. If no feasible action survives the filter, the agent selects a feasible destination node with the highest estimated uncertainty, where ties are broken randomly. Additionally, the model freezes at the current node for a number of time steps (selected to be 10 in the implementation) to ensure that uncertainty is reduced by at least a minimum amount at each node. 
This model architecture, portrayed in  \autoref{fig:model_architecture}, means that the algorithm is designed to overcome major obstacles facing decentralized control algorithms, including the exponential sample complexity with the increase in the number of agents \cite{shi2025breakingcursemultiagencyrobust}, non-stationarity from each agent's perspective \cite{nekoei2023dealingnonstationaritydecentralizedcooperative}, and coordination among decentralized agents \cite{zhang2019decentralizedmultiagentreinforcementlearning}. 

\subsection{Reinforcement Learning Design}
\subsubsection{Reward} In Reinforcement Learning, the reward function $r_{i,k}$ represents the quantity that each agent seeks to maximize over time. Each agent is given its own separate reward function, which its controller uses to modify its behavior. The reward function is the mechanism by which the environment provides behavioral feedback to the agent; it is the means by which decentralized agents can be compelled to achieve a global goal through coordination. In this paper, the reward function for each agent is given by 
\begin{equation}
r_{i,k} \triangleq W_{\texttt{col}} \texttt{col}_i + W_{\texttt{change}}\texttt{change}_i + W_{\texttt{mom}}\texttt{mom}_{i,t}
\end{equation} 
where $A,B,C\in\mathbb{R}$. The terms \texttt{col}$_i$ (collision),
$\texttt{change}_i$, and 
$\texttt{mom}_{i,t}$ (momentum) are defined as 
\begin{equation}
\texttt{col}_i = \begin{cases}
        -3, & p_i\in P_{-i},\\
        0, & \text{otherwise},\\
    \end{cases}
\end{equation}
where $P$ is the set of all current nodes each agent is positioned at and $p_i$ is the position of $\mathcal{A}_i$,
\begin{equation}
\texttt{change}_{i}= \bar{U}-U_i, 
\end{equation} 
where $U_i = \sum_{k=1}^{K} \sum_{m=1}^{M} u^{(e)}_{m,k}$, as in $U_i$ equals the sum of uncertainty over all time-steps and nodes of $\mathcal{M_i}$ for the current episode, and $\bar{U}$ is the average of $U_i$ over the past $W \in\mathbb{Z}$ episodes with $W=10$, and 
\begin{equation}
    \texttt{mom}_{i,t+1} = 
    \begin{cases}
    \texttt{mom}_{i,t}+1,  &u_i>0,\\
    \texttt{mom}_{i,t}-1,  &\text{otherwise},\\
    \end{cases}
\end{equation} 
where $u_i$ represents the uncertainty at the node the agent remains; momentum resets at the start of each episode.
Additionally, the terms $A, B, C$ are each set to $0.05$ to avoid excessively large reward signals, as higher values caused unstable training, based on inspection of the shape of the reward curves during training.
\vspace{-2mm}
\subsection{Modeling with Networks}
The applicability of this algorithm is not limited to scenarios that are immediately conducive to the usage of networks (e.g., power grids). This is because any $n$-dimensional rectangular environment can be divided into discrete components, whose spatial relationships can be represented as edges and whose space may be modeled as a graph with nodes $\mathcal{V}$ representing parts of the space. Given $n$ dimensions of length $d_n$: ${d_1,d_2,...,d_n},d\in\mathbb{R}$ and a desired number of chunks taken from $d_n$, $m_n, m_n\in\mathbb{R},d_n\%m_n=0$, Algorithm \ref{alg:domain_partition} is able to decompose that space into a graph that represents all of the space as nodes and contact between parts of the space as edges.

\begin{algorithm}
\caption{Partitioning an $n$-rectangular region into graph nodes and edges}
\label{alg:domain_partition}
\begin{algorithmic}[1]
    \State \textbf{Input:} $n$-dimensional rectangular space with dimension lengths $d_{1:n}$ 
    \Statex \Comment{A hyperrectangle is given to the algorithm, where it is partitioned into a list of nodes representing equally spaced portions of space in the input region.}
    \State $\text{listOfNodes} \gets [\,]$ 
    \Comment{A list containing each node is initialized}
    \For{$M_1 \text{ in range}(1, \frac{d_1}{m_1})$}
        \For{$M_2 \text{ in range}(1, \frac{d_2}{m_2})$}
            \Statex \hspace{1cm}$\ddots$
            \For{$M_n \text{ in range}(1, \frac{d_n}{m_n})$} \State
            $\texttt{C}_k\triangleq((M_k-1)m_k,M_km_k)), \forall k\in\mathbb{N}_n$
            \State region $\gets \big(\texttt{C}_1,  \texttt{C}_2,..., \texttt{C}_n\big)$ 
            \Statex \Comment{Coordinates at a given portion of a region are assigned to a tuple.}
            \State 
            $\text{node}_{M_1,M_2,\ldots,M_n} \gets$ region
            \Statex \Comment{Each node is iteratively assigned coordinates from the lengths, ensuring the entire region is recorded as a node.}
            \State $\text{listOfNodes.append}(\text{node})$
            \EndFor
        \EndFor
    \EndFor  
    \State $\text{edges} \gets \varnothing$
    \State edge $\gets$ True
    \For{node in listOfNodes}
        \For{$i_1\in\{-1,0,1\}$}
            \Statex \hspace{1cm}$\ddots$
            \For{$i_n\in\{-1,0,1\}$}
                
                \For{coordinate in node.coordinates}
                    \If{index(coordinate) + $i_{1:n} \geq -1$}
                        \State $\text{edge} \gets \text{True}$
                        \State \textbf{continue}
                    \Else
                        \State $\text{edge} \gets \text{False}$
                        \State \textbf{break}
                    \EndIf
                \EndFor
                
                \If{$\text{edge} = \text{True}$}
                    \State $\text{edges} \gets \text{edges} + 
                    (\text{node},\text{node}(i_{1:n}))$
                \EndIf
                
            \EndFor
        \EndFor
    \EndFor
\State \textbf{Output:} listOfNodes, edges 
\Statex \Comment{The algorithm outputs a list of nodes indexed by space, as well as a list of edges connecting the nodes, given by their indices.}
\end{algorithmic}
\end{algorithm}
\section{Numerical Results} 
\label{sec: numerical results}
To test the proposed algorithm, we modeled the problem in a Python simulation and collected data on various performance metrics, including average uncertainty within the system over time and the maximum number of time steps between node visits, to assess its effectiveness. To create an environment that accurately reflected the persistent monitoring problem in networks, we used various software packages, including NetworkX \cite{networkxpaper2008}, PyTorch \cite{paszke2019pytorchimperativestylehighperformance}, and PettingZoo \cite{terry2021pettingzoogymmultiagentreinforcement}. The following experiments were run using the custom simulation software\footnote{Available publicly at \href{https://github.com/theohrogalski/dpmrl}{https://github.com/theohrogalski/dpmrl}}, with the compute comprising an AMD Ryzen 9 CPU and NVIDIA GeForce RTX 4070 Laptop GPU. Training and testing took approximately 24 and 12 hours, respectively.   
This paper includes in Table \ref{tab:comparison} and Table \ref{tab:permutation} the results of two experiments to assess the proposed algorithm: the first, when compared to other versions of the same algorithm, and the second, when compared to variations of itself, suggesting optimality among various possible designs.
\vspace{-1.5mm}
\subsection{Metrics}
Two metrics were collected on a collection of randomly seeded solution architectures:
\begin{itemize}
    \item \textbf{Average uncertainty over time}: The sum of each node's uncertainty, averaged over time using the arithmetic mean, to provide a clearer statistical picture of uncertainty reduction. 
    \item \textbf{Max. time between any target visit}: The maximum number of time steps between an agent visiting a node, capturing the tendency of a policy to neglect nodes by keeping agents in a small area.
\end{itemize}

These metrics capture key aspects of the persistent monitoring problem, providing insight into the relative performance of a given control policy.
\subsection{Experimental Setup}

A number of system parameters, including the number of targets and nodes, are varied to assess the algorithm's effectiveness across different scenarios with four agents. Each network is constructed according to a random process in which the probability of an edge between two nodes is 20\%, and the probability that a node is a target is set to approximately 16\%. Three representative situations are tested: the first with 50 nodes and 8 target nodes; the second with 100 nodes, 100 agents, and 17 target nodes; and the third with 200 nodes, 200 agents, and 32 target nodes, each with 4 agents. The uncertainty at a given node is capped at 100 to reflect the minimal difference in information loss from neglecting a node for 100 or more time steps in a real-world scenario. Each agent is trained for 100 episodes, with 500 moves per episode, a number of steps that reflects a finite-time monitoring scenario such as information gathering in a limited time span over a certain area.
\begin{figure}
\includegraphics[width=0.5\textwidth]{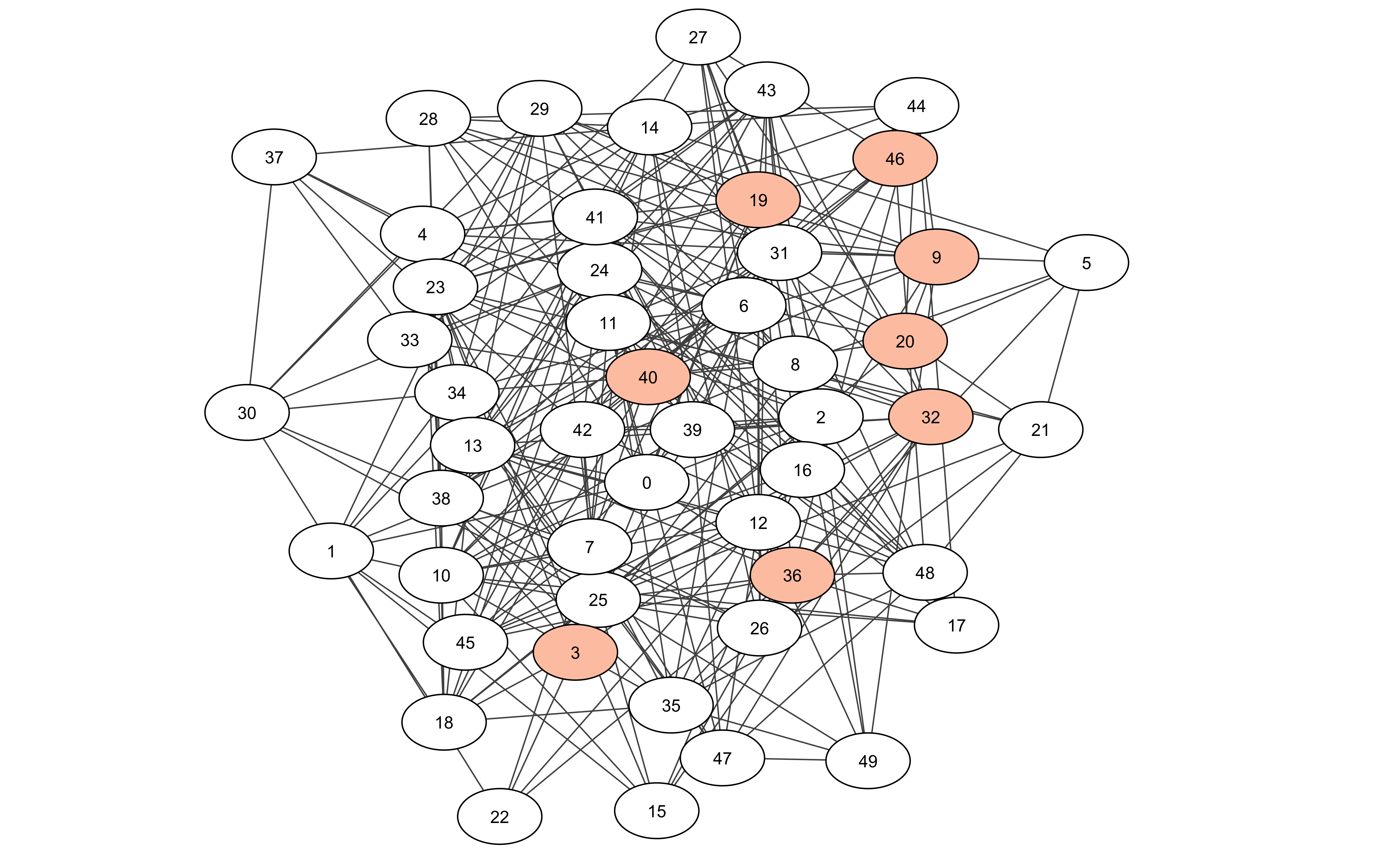}
\caption{Network with 50 nodes, eight targets (highlighted in peach) and connections between 20\% of nodes.}
\label{fig:network50}
\end{figure}

The number of agents is held constant to reduce computational complexity and assess a realistically sized team's ability to tackle increasingly large-scale problems. Figure \ref{fig:network50} illustrates the environment in which each agent reasons. Each graph keeps the same structure during training and testing.

\begin{figure*}[!t]
\centering
\includegraphics[width=0.75\textwidth]{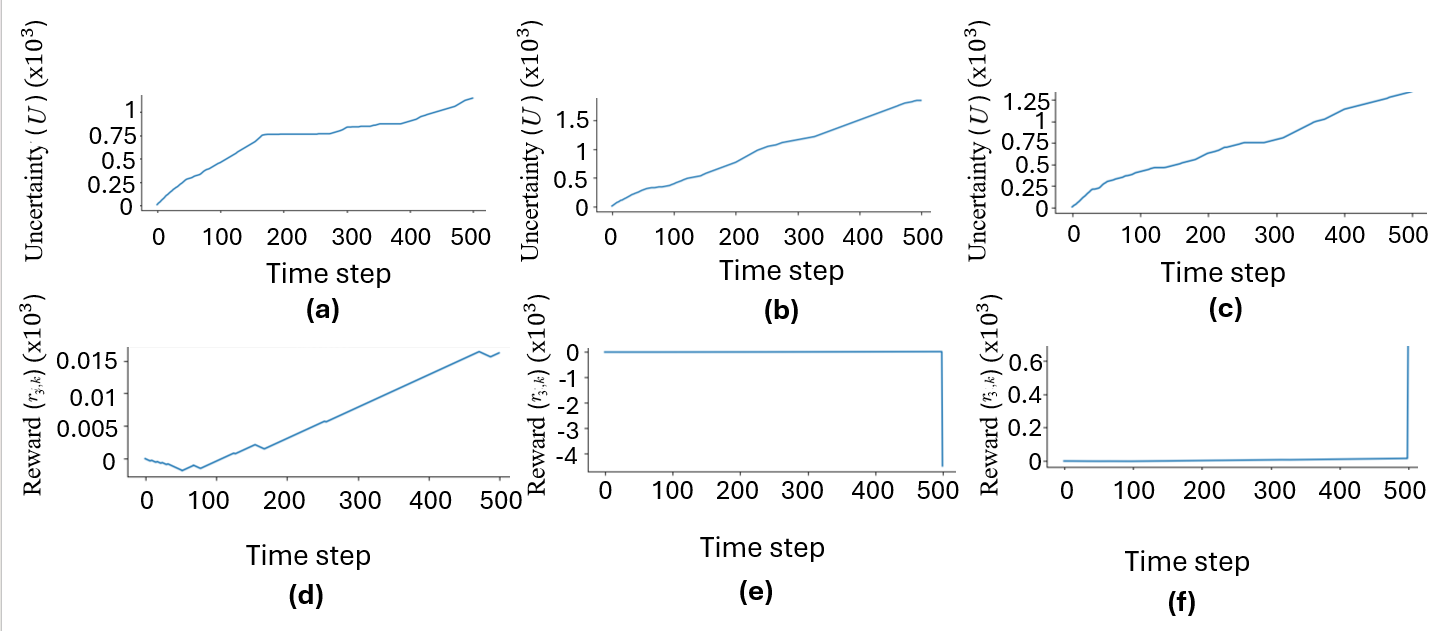}
\vspace{-1mm}
\caption{
Various reward (above) and reward (below) curves for different parts of the training process, with curves (a) and (e) representing episode one, curves (b) and (d) representing episode seven, and curves (c) and (f) representing episode 15, demonstrating the process of moving away from a random control policy to a more effective one over time
}
\vspace{-5mm}
\label{fig:reward_and_uncertainty_curves}
\end{figure*}

\begin{figure}[!t]
\centering
\includegraphics[width=0.5\textwidth]{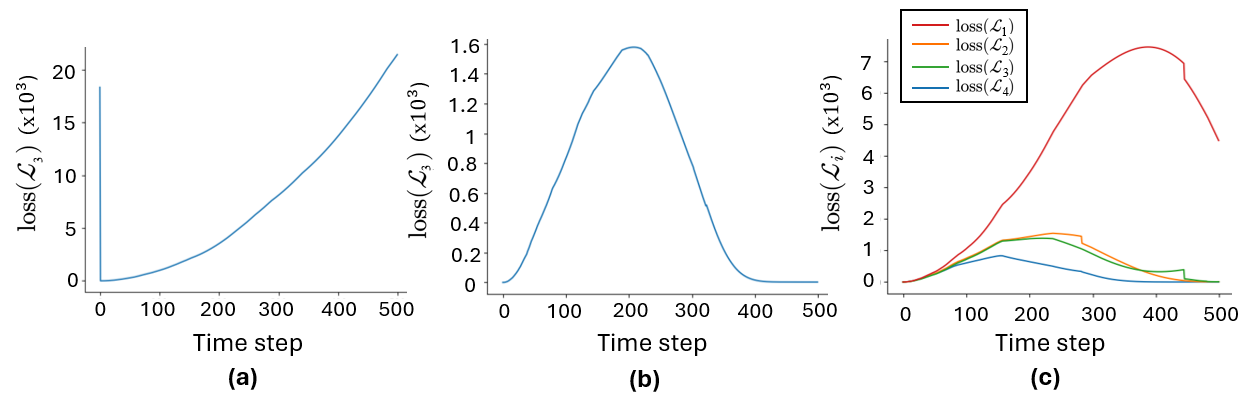}
\vspace{-1mm}
\caption{An exponentially increasing pattern for $\mathcal{A}_3$ for one episode (a), a pattern from $\mathcal{A}_3$ demonstrating a low-loss pattern for one episode (b) \& state estimator loss at one early episode  with the proposed model architecture on a graph with 100 nodes for four agents (c)}
\label{fig:state_est_curves}

\end{figure}
Hyperparameters for training and testing include a learning rate of $\frac{1}{1000}$ for the optimizer (Adam). Uncertainty curves for one episode alongside the corresponding reward for $\mathcal{A}_3$, areshown in Figure \ref{fig:reward_and_uncertainty_curves}. The loss of the state estimator for four agents over a single episode is shown in Figure \ref{fig:state_est_curves}.
The data in the following tables were collected as follows: ten runs of 500 steps of the model in the environment were collected for three seeds and three graphs, giving 90 data points for each model (excepting the centralized model, which was run for one seed due to computational constraints leading to around a fivefold increase in training time, possibly due to the relative sparsity of the decentralized mental maps).

\subsection{Experiment One: Performance Benchmarking}
We compare against baseline algorithms with various capabilities, including an omniscient centralized algorithm (grazing), a random action-selection policy (random policy), and a simple policy of selecting a target to stay at until the run completes (sit-on-nodes policy). A centralized algorithm with a shared mental map $\mathcal{M}$, shared state estimator $\mathcal{L}$, and shared observation processing network similar to \cite{zhangcompass} is also tested, allowing for a strong comparison between the effect of full decentralization on the performance of a given algorithm. The implementation in this paper differs from \cite{zhangcompass} in that it does not incorporate state history, spatio-temporal attention, or other components that rely on the environment presented in that work. Each algorithm represents a different approach to persistent monitoring, as they are either fully centralized/omniscient algorithms or simple policies.

\begin{table}[t]
\centering
\caption{Simulation Performance Comparison}
\vspace{-5mm}
\begin{threeparttable}
\label{tab:comparison}
\renewcommand{\arraystretch}{1.5}
\setlength{\tabcolsep}{5pt}
\footnotesize
\begin{tabular}{p{2.25cm}cccc}
\toprule
\textbf{Algorithm} &
\textbf{\shortstack{Avg. \\ Uncertainty}} &
\textbf{\shortstack{Std. Dev. \\ (\textit{Uncertainty})}} &
\textbf{\shortstack{Avg. \\ Neglect \\ Time}} &
\textbf{\shortstack{Std. \\ Dev. \\ (\textit{Neglect})}} \\
\midrule
Proposed Algorithm & 1736.3 & 895 & 1176.6 & 561.6 \\ \hline

Grazing & \textbf{1128.1} & 1078.4 & \textbf{276.7} & \textbf{195.2} \\ \hline

Random & 3006 & 1147.9 & 5001\tnote{*} & 0  \\ \hline

Sit on Nodes & 3006 & 1147.9  & 5001\tnote{*} & 0 \\ \hline

Centralized Algorithm & 2356.7 & 1120 & 322 & 213 \\ 
\bottomrule
\end{tabular}
 \begin{tablenotes}
      \small
      \item[*] \textit{Value reached saturation} 
    \end{tablenotes}
    \end{threeparttable}
    \vspace{-5mm}
\end{table}



From Table \ref{tab:comparison}, the average uncertainty of the grazing algorithm is the best (lowest), while the proposed and centralized algorithms have average uncertainties 53.9\% and 108.9\% higher than Grazing, respectively. However, because the grazing algorithm assumes perfect knowledge of the state, it is impractical for real-world applications, implying that the proposed algorithm may perform better in practice.

The random and sit-on-nodes algorithms fail to perform, resulting in the worst possible value (saturation) occurring in both cases, due to the random algorithm's tendency to move to a connected node rather than stay at the current node. 

The centralized and Grazing algorithms reduce the mean maximum inter-visit time by 72.6\% and 76.5\% respectively relative to the proposed algorithm, suggesting that the decentralized algorithm focuses on optimizing a small local environment rather than exploring the global environment. 

The less-complex sit-on-nodes and random algorithms both performed extremely poorly on this task, achieving the same maximum score. This implies that they were not able to reduce the uncertainty of many, if any, nodes.

\subsection{Experiment Two: Variational and Ablative Analysis}

Though the algorithm performs within the same region as algorithms that contain unrealistic assumptions or exhibit minimal performance (grazing and random), it may be that certain components are unnecessary, or that a slight modification to the algorithm may improve performance. Given that, a variational and ablative study was conducted to assess unnecessary components and potential improvements to the algorithm. These removals and additions include the removals of the DCBF heuristic and the state estimator, a modified DCBF-inspired heuristic that prevents collisions between agents (occupation of the same node at the same time) and the addition of numerous multi- and single-headed attention layers to process the data. 
A single-factor ANOVA on the data yielded a p-value of 0.012 and an F statistic exceeding the critical threshold, indicating a statistically significant result. This suggests that the data each model produces come from substantively different distributions.

\begin{table}[t]
\centering
\caption{Ablation and Variant Analysis}
\vspace{-5mm}
\begin{threeparttable}
\label{tab:permutation}
\renewcommand{\arraystretch}{1.5}
\setlength{\tabcolsep}{5pt}
\footnotesize
\begin{tabular}{p{2.25cm}cccc}
\toprule
\textbf{Algorithm} &
\textbf{\shortstack{Avg. \\ Uncertainty}} &
\textbf{\shortstack{Std. Dev. \\ (\textit{Uncertainty})}} &
\textbf{\shortstack{Avg. \\ Neglect \\ Time}} &
\textbf{\shortstack{Std. \\ Dev. \\ (\textit{Neglect})}} \\
\midrule
Proposed Algorithm & 1736.3 & 895 & 1176.6 & 561.6 \\ \hline

No DCBF & 3006.0 & 994 & 10001 \tnote{*} & 0 \\ \hline

No State Est. & 1770.0 & 831 & 1241.3 & 667.4\\ \hline

No Collisions & 1750.4 & 847 & 1139.3 & 564.3 \\ \hline

Extra Attention & \textbf{1734.6} & \textbf{830} & \textbf{1103.1} & \textbf{439.5} \\ 
\bottomrule
\end{tabular}
 \begin{tablenotes}
      \small
      \item[*] \textit{Value reached saturation} 
    \end{tablenotes}
    \end{threeparttable}
    \vspace{-5mm}
\end{table}

Table \ref{tab:permutation} demonstrates that while the uncertainty performance for the algorithm with extra attention layers is high, it is less than 1\% better performing than the proposed algorithm. However, the extra attention algorithm requires an additional single-head attention and graph transform layer (5 layers and 5$\times$5, respectively, with 1 head each) to function.

For both metrics collected, the No Collisions algorithm performed better and had a similar or lower standard deviation, suggesting that further exploration of it is justified.  

The No Collisions DCBF and No State Estimator algorithms both performed about 1\% less effectively than the proposed algorithm, suggesting potential use cases in situations where collisions are unacceptable or compute resources are limited.

\section{Conclusion} 
\label{sec: conclusion}

This paper presents a fully decentralized, safety-aware MARL algorithm for coordinating a team of non-communicating agents over a network. Incorporating deep reinforcement learning, control barrier functions, and full decentralization enabled strong performance in the persistent monitoring problem. Additionally, this paper elaborates on the importance of problem selection for key algorithms, drawing on prior research \cite{akella2025fundamentallimitationsdecentralizedlearnable}, and demonstrates the potential for fully decentralized solutions to address problems where local sub-objectives are key to global solutions. 
The performance of the proposed algorithm exceeds its centralized counterpart in terms of average uncertainty, indicating that its applicability to real-world drone- and robot-based persistent monitoring is plausible. 
Future directions for this research include identifying additional applications for fully decentralized algorithms, establishing stronger principles for the convergence of fully decentralized systems, and further investigating the impact of safety on control algorithms. 

\bibliographystyle{IEEEtran}
\bibliography{ref} 
\end{document}